\documentstyle{article}
\begin{document}
\hfill{UM-P-99/16}
\begin{center}

{\LARGE \bf Mass Bounds for Flavour Mixing Bileptons \\}
\vspace{10mm}
{\bf M.B.Tully 
\footnote{E-mail: mbt@physics.unimelb.edu.au}
and G.C.Joshi
\footnote{E-mail: joshi@bradman.ph.unimelb.edu.au}
}\\
\vspace{10mm}
{\sl Research Centre for High Energy Physics,\\
School of Physics, 
University of Melbourne,\\
Parkville, Victoria 3052, Australia.}
\end{center}
\vspace{30mm}
\begin{abstract}
Mass bounds for doubly-charged bilepton gauge bosons are
derived from constraints on fermion pair production
at LEP      
and lepton-flavour violating charged lepton decays.
The limit obtained of 740 GeV for the doubly-charged
bilepton does not depend on the assumption that the
bilepton coupling is 
flavour-diagonal, unlike other bounds which have
been given in the literature.

\end{abstract}
\newpage 
\section{Introduction}
Bileptons, gauge bosons carrying lepton number
$L=2$, \cite{Weinberg} arise in a class of models in which the
Standard Model gauge group 
$SU(3)_c \times SU(2)_L \times U(1)_Y$
is extended to 
$SU(3)_c \times SU(3)_L \times U(1)_X$, known
as 331 Models. \cite{B0}-\cite{A57}
Recently, a new lower limit on the mass of the
doubly-charged bilepton of 850 GeV has been
obtained from bounds on the conversion of
muonium ($\mu^-e^+$) to antimuonium ($\mu^+e^-$).
\cite{Willmann}
However, it has been noted \cite{Das,Pleitez} 
that this limit
relies on the assumption that the matrix $V_Y$
which couples bileptons to ordinary leptons is
at least approximately flavour diagonal, since
the predicted conversion rate is dependent on the product
 $\left(V_Y^{11}V_Y^{22} \right)$. 
It is therefore important to search for mass
limits that allow for the possibility of
more general coupling.

In particular, as will be discussed below, 
the scenario in which
$V_Y^{11} \simeq V_Y^{22} \simeq 0$ does not
appear to be in conflict with experimental
data, and in this situation the muonium
to antimuonium conversion rate would be zero,
and so the 850 GeV mass limit would not apply.

The method to be employed here uses the fact
that doubly-charged bileptons can
contribute to 
electron - positron
scattering
by $u$-channel exchange, as
first pointed out by Frampton and Ng \cite{C51},
and also considered by Cuypers and Davidson \cite{A26}.
Recently, new bounds on 
the mass scale of any exotic contributions
to $e^+e^- \rightarrow e^+e^-$
have been obtained by the OPAL collaboration at LEP
\cite{OPAL}
which can thus be used to constrain the
bilepton mass.
Further, new limits on exotic
contributions to 
$e^+e^- \rightarrow \mu^+\mu^-$
and $e^+e^- \rightarrow \tau^+\tau^-$
have also been obtained.
These three processes have 
dependence on the matrix elements $V_Y^{11}$,
$V_Y^{12}$ and
$V_Y^{13}$
 respectively. Since $V_Y$ is unitary,
an absolute bound on the bilepton mass
can then be given, regardless of the nature
of the coupling.
In this manner, we find a lower bound on the
doubly charged bilepton mass of 510 GeV.

This limit can be increased if data from 
lepton-flavour violating charged lepton decays
is also taken into account, as these
limits strongly constrain the form of $V_Y$.
By combining these limits with the pair production
data, the bilepton mass bound is 
able to be increased
to  740 GeV.

\section{Fermion Pair Production}
For 
$M_Y \gg \sqrt{s}$
as is the case here, the process 
$e^+e^- \rightarrow f \overline{f}$
may be treated by the four-fermion contact
interaction formalism.
In the usual parameterisation \cite{Peshkin}, 
the effective
Lagrangian 
is given by
\begin{equation}
{\cal L}^{contact} = 
\frac{g^2}{(1+ \delta )\Lambda^2} 
\sum_{i,j=L,R}
\eta_{ij} 
\left[ \overline{e}_i \gamma^\mu e_i \right]
\left[ \overline{f}_j \gamma_\mu f_j \right]
\label{eq:4fc}
\end{equation}
(where the symmetry factor $\delta=1$ for $f=e$ 
and 0 otherwise).
The coupling $g$ in equation \ref{eq:4fc} is conventionally  
set to
$g^2 / 4 \pi =1$.

A bilepton vertex is of the form
\begin{equation}
\frac{g_{3l}}{\sqrt{2}} \overline{l}^c_i 
\gamma_\mu \gamma_L V^{ij} l_j Y^{++}
\end{equation}
writing $l= \left( e,\mu, \tau \right)$, leading to
a four-fermion interaction: 
\begin{eqnarray}
& &
V^{ij}V^{kl \star} \frac{g_{3l}^2} {2M_Y^2}
\overline{l}^c_i \gamma^\mu \gamma_L  l_j 
\overline{l}_l \gamma_\mu \gamma_L  l^c_k
\\
 & = & 
V^{ij}V^{kl \star} \frac{g_{3l}^2} {2M_Y^2}
\overline{l}^c_i \gamma^\mu \gamma_L  l^c_k 
\overline{l}_l \gamma_\mu \gamma_L  l_j
\\
& = & - 
V^{ij}V^{kl \star} \frac{g_{3l}^2} {2M_Y^2}
\overline{l}_k \gamma^\mu \gamma_R  l_i 
\overline{l}_l \gamma_\mu \gamma_L  l_j
\end{eqnarray}
Expressing this in the form of equation \ref{eq:4fc}
gives $\eta_{RL}=\eta_{LR}=+1$ (the negative sign from 
the Fierz transformation cancelling with an overall
negative sign due to the relative ordering of the
fermion fields \cite{C51}) with couplings given
in the three cases by:
\begin{equation}
\begin{array}{ccc}
e^+e^- \rightarrow e^+e^- &
\eta_{LR} &
\frac{g^2}{\Lambda^2} = 
\frac{g_{3l}^2}{2M_Y^2}|V_Y^{11}|^2\\
 &
\eta_{RL} &
\frac{g^2}{\Lambda^2} = 
\frac{g_{3l}^2}{2M_Y^2}|V_Y^{11}|^2\\
\\
e^+e^- \rightarrow \mu^+ \mu^- &
\eta_{LR} &
\frac{g^2}{\Lambda^2} = 
\frac{g_{3l}^2}{2M_Y^2}2|V_Y^{12}V_Y^{21}|\\
\\
 &
\eta_{RL} &
\frac{g^2}{\Lambda^2} = 
\frac{g_{3l}^2}{2M_Y^2}\left( |V_Y^{12}|^2+|V_Y^{21}|^2 \right)\\
\\

e^+e^- \rightarrow \tau^+ \tau^- &
\eta_{LR} &
\frac{g^2}{\Lambda^2} = 
\frac{g_{3l}^2}{2M_Y^2}2|V_Y^{13}V_Y^{31}|\\
\\
&
\eta_{RL} &
\frac{g^2}{\Lambda^2} =
\frac{g_{3l}^2}{2M_Y^2}\left( |V_Y^{13}|^2+|V_Y^{31}|^2 \right)\\

\end{array}
\end{equation}
For convenience 
the approximation will now be made that 
$|V_Y^{ij}|=|V_Y^{ji}|$,
experimentally justified by  the charged lepton decay
limits discussed below. The combined limit $\Lambda_{LR+RL}$
can then be used rather than the separate limits 
$\Lambda_{LR} = \Lambda_{RL}$.
Specialising to the 331 Model, 
the bilepton coupling $g_{3l}$ will be set 
to $g=e/\sin \theta_W$.

The limits then obtained are shown in table \ref{table:mass}.

\begin{table}[h]
\[
\begin{array}{|c|c|c|}
\hline
$Process$ & $Mass Scale$ & $Bilepton Mass Limit$ \\
\hline
e^+e^- \rightarrow e^+e^-         & 
\Lambda > 6.2 {\rm TeV} &
M_Y > |V^{11}| 770 {\rm GeV} \label{eq:mlim1}\\
e^+e^- \rightarrow \mu^+\mu^-     &
\Lambda > 4.9 {\rm TeV} &
M_Y > |V^{12}| 860 {\rm GeV} \label{eq:mlim2} \\
e^+e^- \rightarrow \tau^+\tau^-   &
\Lambda > 6.3 {\rm TeV} &
M_Y > |V^{13}| 1110 {\rm GeV} \\
\hline
\end{array} 
\]
\caption{Limits on mass scale of new contact 
interaction \cite{OPAL} and corresponding 
bilepton mass limit (95\% C.L.)}
\label{table:mass}
\end{table}

Consideration of this data alone gives
a lower limit on the bilepton mass of 
$M_Y > 512 {\rm GeV}$
with $|V^{11}|=0.66$, $|V^{12}|=0.59$ and $|V^{13}|=0.46$,
however this limit can be made more restrictive 
by including additional experimental data.

\section{Charged Lepton Decay Limits}
Information on the form of $V_Y$ can be deduced 
from limits on the exotic decays of $\mu$ and
$\tau$ into three charged leptons, that is
$\mu^- \rightarrow e^+e^-e^-$ and 
$\tau^- \rightarrow l_1^+l_2^-l_3^-$,
where $l_i=e,\mu$. \cite{A57,A26} 
Such decays could be mediated
by bileptons with appropriate couplings,
however current experimental limits 
on the branching ratio for these decays are
$1 \times 10^{-12}$ for muon decays, \cite{Bellgardt} and
of the order $10^{-6}$ for the various tau decays. \cite{Bliss}

These limits clearly are compatible with low-mass
bileptons only if the coupling matrix is either
almost diagonal, that is:
\begin{equation}
|V_Y^{ij}| \simeq 
\left( \begin{array}{ccc}
1 & 0 & 0 \\
0 & 1 & 0 \\
0 & 0 & 1 
\end{array}
\right)
\label{eq:vyform1}
\end{equation}
or of the form
\begin{equation}
|V_Y^{ij}| \simeq 
\left( \begin{array}{ccc}
0 & 1 & 0 \\
1 & 0 & 0 \\
0 & 0 & 1 
\end{array}
\right)
\label{eq:vyform2}
\end{equation}
in which mixing between the first
two generations is maximal.
Of course though, $V_Y$ becomes less restricted
as the value of $M_Y$ is increased.
(An interesting question we do not consider
here is that of the naturalness of these
forms for $V_Y$).

In particular, limits on the decays
$ \mu^-  \rightarrow  e^+ e^- e^-$, 
$\tau^-  \rightarrow  e^+ e^- e^-$ and 
$\tau^-  \rightarrow  e^+ \mu^- e^-$ 
respectively
lead to the following constraints:
\begin{eqnarray}
\left( \frac{M_W}{M_Y} \right)^4
\left( |V_{12}|^2+|V_{21}|^2 \right) |V_{11}|^2 
& < & 1.0 \times 10^{-12} \label{eq:declim1}\\
\left( \frac{M_W}{M_Y} \right)^4
\left( |V_{13}|^2+|V_{31}|^2 \right) |V_{11}|^2 
& < & 1.7 \times 10^{-5} \label{eq:declim2}\\
\left( \frac{M_W}{M_Y} \right)^4
\left( |V_{13}|^2+|V_{31}|^2 \right) 
\left( |V_{12}|^2 + |V_{21}|^2 \right) 
& < & 1.0 \times 10^{-5} \label{eq:declim3}
\end{eqnarray}

A bilepton mass limit can then be obtained for
each of the two possible forms of $V_Y$
(equations \ref{eq:vyform1} and \ref{eq:vyform2})
by requiring that the limits from table
\ref{table:mass} and from equations
\ref{eq:declim1}-\ref{eq:declim3}
are all satisfied, consistent with $V_Y$
being unitary.

For flavour-diagonal coupling (equation \ref{eq:vyform1}) 
the limit obtained in this manner is
\begin{equation}
M_{Y^{--}} > 740 {\rm GeV}
\end{equation}
with $|V_Y^{11}|=0.97$,
while for the case of maximal mixing between the
first and second generations (equation \ref{eq:vyform2})
the limit is
\begin{equation}
M_{Y^{--}} > 860 {\rm GeV}
\end{equation}
with $|V_Y^{12}|=0.99$.

\section{Conclusion}
A lower bound on the mass of the doubly-charged
bilepton has been obtained of
\begin{equation}
M_{Y^{--}} > 740 {\rm GeV}
\end{equation}
from consideration of experimental limits on
fermion pair production in electron-positron collisions
and lepton-flavour
violating charged lepton decays.
While this limit is less stringent than
that obtained from muonium-antimuonium conversion,
it is more general, as it does not depend on any
assumptions about the form of the
bilepton mixing matrix.

This value may also be compared with the mass bound
on the singly charged bilepton of 440 GeV \cite{Tully}
obtained from consideration of muon decay parameters,
and which is also independent of the form of $V_Y$.

\section*{Acknowledgements}
The authors would like to thank Robert Foot for
a
helpful discussion.

\end{document}